\documentclass{icrc2009}
\usepackage{amssymb}
\usepackage{graphicx}   
\usepackage[caption=false]{caption}    
\usepackage[font=footnotesize]{subfig} 
\usepackage{fixltx2e}
\usepackage{url}

\newcommand{\shorttitle}[1]%
{\markboth{Proceedings of the 31\MakeLowercase{$^{st}$} ICRC, {\L}\'{o}d\'{z} 2009}{#1} }
\newcommand{\etal}{\MakeLowercase{\textit{et al.}}} 

\begin{document}
\title{Development of neutrino initiated cascades at mid and high altitudes in the atmosphere}

\author{\IEEEauthorblockN{A.D. Supanitsky\IEEEauthorrefmark{1},
			  G. Medina-Tanco\IEEEauthorrefmark{1},
                          K. Asano\IEEEauthorrefmark{2},
                          D. Cline\IEEEauthorrefmark{4},
                          T. Ebisuzaki\IEEEauthorrefmark{5},\\
                          S. Inoue\IEEEauthorrefmark{6},
                          P. Lipari\IEEEauthorrefmark{7},
                          A. Santangelo\IEEEauthorrefmark{8},
                          K. Shinozaki\IEEEauthorrefmark{5},
                          G. Sigl\IEEEauthorrefmark{9},
                          Y. Takahashi\IEEEauthorrefmark{10} \\ and 
                          M. Teshima\IEEEauthorrefmark{11} for the JEM-EUSO Collaboration
} \\

\IEEEauthorblockA{\IEEEauthorrefmark{1} Departamento de F\'isica de Altas Energ\'ias, Instituto de Ciencias 
Nucleares, Universidad Nacional Aut\'onoma \\ de M\'exico, A. P. 70-543, 04510, M\'exico, D. F., M\'exico.}
\IEEEauthorblockA{\IEEEauthorrefmark{2} Interactive Research Center of Science, Graduate School of Science, 
Tokyo Institute of Technology, \\ 2-12-1 Ookayama Meguro-ku Tokyo 152-8550, Japan.}
\IEEEauthorblockA{\IEEEauthorrefmark{4} Department of Physics and Astronomy, University of California, Los Angles, USA.}
\IEEEauthorblockA{\IEEEauthorrefmark{5} RIKEN Advanced Science Institute, Japan.}
\IEEEauthorblockA{\IEEEauthorrefmark{6} Dept. of Physics, Kyoto University, Kyoto 606-8502, Japan.}
\IEEEauthorblockA{\IEEEauthorrefmark{7} INFN-Roma La Sapienza, I-00185 Roma, Italy.}
\IEEEauthorblockA{\IEEEauthorrefmark{8} Institute fuer Astronomie und Astrophysik Kepler Center for Astro and 
Particle Physics\\ Eberhard Karls University Tuebingen Germany.}
\IEEEauthorblockA{\IEEEauthorrefmark{9} Institut theoretische Physik Universitaet Hamburg Luruper Chaussee 149 
D-22761 Hamburg, Germany.}
\IEEEauthorblockA{\IEEEauthorrefmark{10} Dept. of Physics, The University of Alabama in Huntsville, Huntsville, 
AL35899, USA.}
\IEEEauthorblockA{\IEEEauthorrefmark{11} Max-Planck-Institut f\"ur Physik, F\"ohringer Ring 6, D-80805 M\"unchen, 
Germany.}

}

\shorttitle{A.~D.~Supanitsky \etal Neutrino initiated cascades at mid and high altitudes}
\maketitle

\begin{abstract}

Neutrinos are a very promising messenger at tens of EeV and above. They can be produced by several channels, 
namely as by products of hadronic interactions at the sources, as the main products of the decay of super 
massive particles and, in a guaranteed way, as the result of the propagation of UHECR through the bath of 
microwave relic photons. A new era of very large exposure space observatories, of which the JEM-EUSO mission 
is a prime example, is on the horizon and, with it, it is even larger the possibility of astrophysical neutrino 
detection at the highest energies. In the present work we use a combination of the PYTHIA interaction code with 
the CONEX shower simulation package in order to produce fast one-dimensional simulations of neutrino initiated 
showers in air. We make a detail study of the structure of the corresponding longitudinal profiles, but focus 
our physical analysis mainly on the development of showers at mid and high altitudes, where they can be an 
interesting target for space fluorescence observatories.

\end{abstract}


\section{Introduction}

The neutrino flux carries very important astrophysical information. A high energy neutrino flux 
is expected as a by-product of the interactions of cosmic ray hadrons at the sources \cite{CentaurusA:08}. 
They can also be produced during the propagation of cosmic rays in the intergalactic medium 
\cite{Berezinsky:69} and as the main product of the decay of superheavy relic particles 
\cite{Aloisio:04,Battacharjee:00}.

JEM-EUSO \cite{Ebisuzaki:09} with its $10^{12}$ tn of atmospheric target volume has the real possibility 
of observing high energy neutrinos and make important contributions to the understanding of UHECR production 
and propagation \cite{Gustavo:09,Palomares:05}. Source distributions rapidly evolving with redshift would be 
particularly favorable by increasing the cosmogenic neutrino flux at highest energies \cite{Berezinsky:69}. 
A thorough understanding of neutrino deep inelastic scattering, as well as the evolution of longitudinal 
profiles of atmospheric neutrino showers, are extremely important in order to take advantage of the full 
potential of the experiment. Conversely, besides the obvious astrophysical value, the properties of just a 
few observed showers can also give valuable information on the physics governing high energy neutrino-nucleon 
interactions. The objective of this work is to present the first part of an ongoing effort in that direction.

\section{Neutrino nucleon interaction}

High energy neutrinos that propagate in the Earth atmosphere can interact with protons and neutrons of 
the air molecules. There are two possible channels for this interaction, charge and neutral current. The 
major uncertainty on the differential cross section at the energies considered comes from the unknown 
behavior of the parton distribution functions (PDFs) at very small values of the parton momentum 
fraction $x$.

\begin{figure}[!t]
\centering
\includegraphics[width=3.2in]{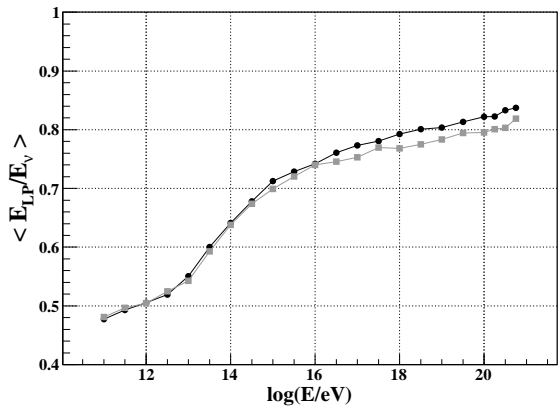}
\caption{Average energy fraction taken by the electron in a proton-electron neutrino charge current
interaction obtained from PYTHIA with CTEQ6 (black circles) and GJR08 (gray squares) sets of PDFs.}
\label{Efract}
\end{figure}

The simulation of neutrino nucleon interaction is performed by using the PYTHIA code \cite{pythia}. 
The parton distribution library LHAPDF \cite{Lhapdf} is linked with PYTHIA in order to be able to use 
different extrapolations of the PDFs. In this work the CTEQ6 \cite{cteq6} and GJR08 \cite{gjr08} PDF 
sets are considered. Fig. \ref{Efract} shows the energy fraction carried by the leading particle as 
a function of the incident neutrino energy for the charge current interaction of an electron neutrino 
with a proton for both sets of PDFs considered. In both cases the energy fraction increases steadily 
with the incoming neutrino energy. The difference between both PDFs increases up to a maximum of a few 
percent at the highest energies.

\begin{figure}[!t]
\centering
\includegraphics[width=3.2in]{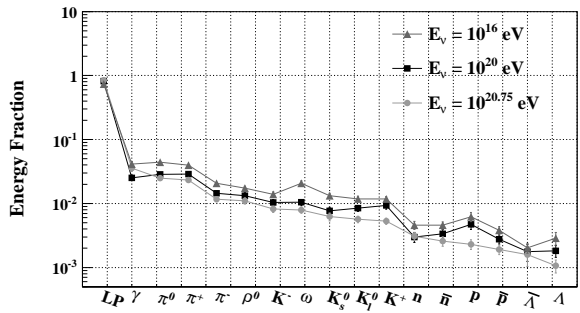}
\caption{Average energy fraction for the most important particles recognized by CONEX produced in a 
charge current interaction of an electron neutrino with a proton for three different neutrino energies.}
\label{EFracCX}
\end{figure}

Besides the leading particle, different types of secondaries are generated as a result of the interaction.
In particular, in this work  we are interested in the ones recognized by CONEX \cite{conex} code which is 
used to simulate the neutrino showers. Fig. \ref{EFracCX} shows the energy fraction taken by the most 
relevant particles recognized by CONEX for three different electron neutrino energies, obtained as a 
result of the charge current interaction with a proton. It can be seen that the smaller the neutrino 
energy the larger the energy fraction taken by the secondary particles. 

\section{Neutrino showers}

\begin{figure}[!t]
\centering
\includegraphics[width=3.2in]{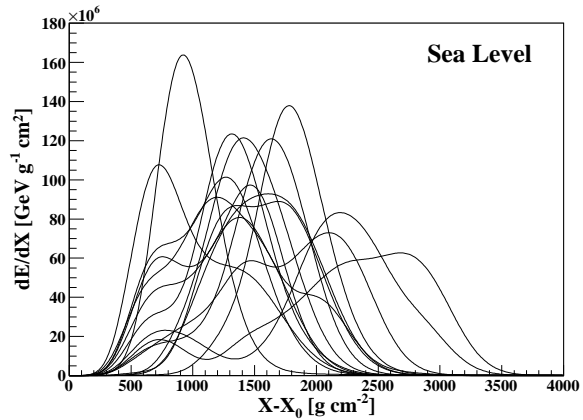}
\caption{Horizontal electron neutrino showers at sea level of $E_\nu = 10^{20}$ eV, injected at 
$X_0=36500$ gcm$^{-2}$, in a point contained on the vertical axis of the FOV.}
\label{NuSh}
\end{figure}

The particles produced in a neutrino-nucleon interaction are injected in CONEX with QGSJET-II \cite{qgII} 
producing extensive air showers. Because the mean free path of neutrinos propagating in the atmosphere is 
very small, they can interact very deeply, after traversing a large amount of matter. Fig. \ref{NuSh} 
shows the energy deposit as a function of $X-X_{0}$, where $X_0$ corresponds to the injection point, for 
horizontal electron neutrino showers at sea level of $E_\nu=10^{20}$ eV, injected at $X_0=36500$ g cm$^{-2}$ 
(maximum slant depth for a completely horizontal shower). Note that the interaction point is situated on 
the vertical axis of the JEM-EUSO field of view (FOV).     
Fig. \ref{NuSh} shows very broad profiles which may present several peaks and large fluctuations. This 
behavior is due to the Landau Pomeranchuk Migdal (LPM) effect, which is very important inside dense regions 
of the atmosphere and for electromagnetic particles, electrons in this case, which take about 80\% of the 
parent neutrino energy.  

\begin{figure}[!t]
\centering
\includegraphics[width=3.2in]{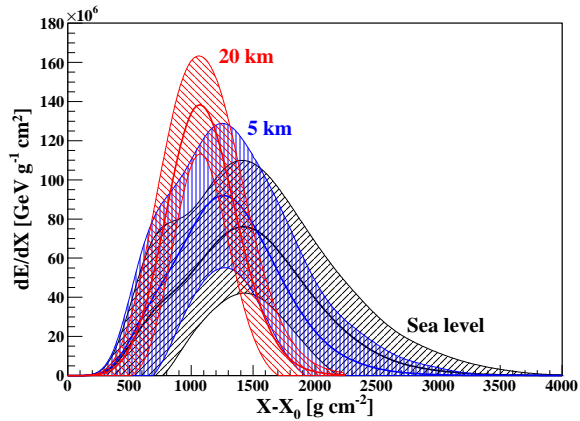}
\caption{Mean value and one sigma regions of the longitudinal profile corresponding to horizontal electron 
neutrino showers of $E_\nu = 10^{20}$ eV for different altitudes. The interaction point is contained on the 
vertical axis of the JEM-EUSO telescope.}
\label{NuShProf}
\end{figure}

An orbital detector like JEM-EUSO can also detect horizontal showers that do not hit the ground. In particular, 
horizontal neutrinos can interact at higher altitudes producing a shower observable by the detector. Fig.  
\ref{NuShProf} shows the mean value and one sigma confidence level regions for the longitudinal profiles for 
horizontal electron neutrino air showers of $E_\nu = 10^{20}$ eV, injected at different altitudes in points 
contained on the vertical axis of the cone of the FOV.   
It can be seen that as the altitude increases the fluctuations are reduced and, on average, the profiles
become thinner. This is due to the fact that the LPM effect become progressively less important with increasing 
altitude because of the decrease in atmospheric density. Note that just showers up to 20 km of altitude are 
considered because, at higher altitudes, the grammage of the FOV is not enough to contain the whole profiles.    

\begin{figure}[!t]
\centering
\includegraphics[width=3.1in]{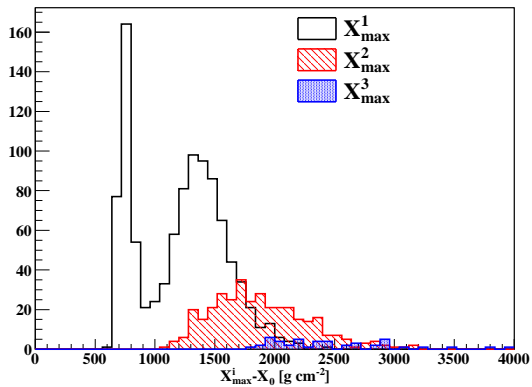}
\caption{Distribution of the position of the maxima for electron neutrino showers of $E_\nu = 10^{20}$ eV
at sea level.}
\label{XmaxI}
\end{figure}

As already mentioned, at smaller altitude the longitudinal profiles present a complicated structure (see Fig. 
\ref{NuSh}). In particular, the showers present multiple peaks. Fig. \ref{XmaxI} shows the distribution functions 
of the position of each maximum. $X_{max}^1$ corresponds to the position of the first maximum counted from the 
start of the shower, $X_{max}^2$ is the second one and so on.   
Note that the distribution function of $X_{max}^1$ is bi-valued and its first peak is located at 
$\sim 800$ g cm$^{-2}$, while the second one is at $\sim 1500$ g cm$^{-2}$. The first peak corresponds to 
the development of the hadronic component of the electron neutrino cascade, whereas the second one reflects 
the electromagnetic portion of the shower.  

\begin{figure}[!t]
\centering
\includegraphics[width=3.3in]{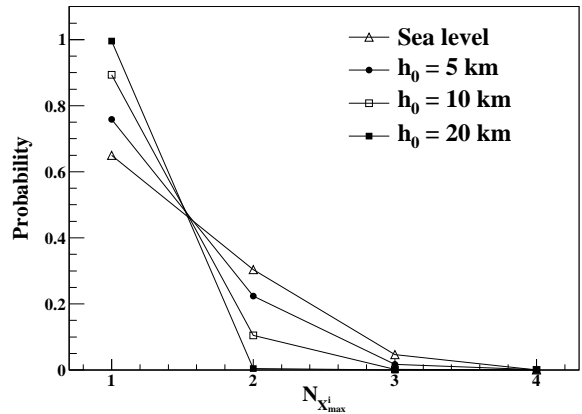}
\caption{Probability of an electron neutrino shower to have $N_{X_{max}^i}$ maxima for $E_\nu = 10^{20}$ eV
for horizontal showers at different altitudes.}
\label{NXmaxI}
\end{figure}

This complicated structure simplifies as the altitude increases. Fig. \ref{NXmaxI} shows the probability
of finding a profile with a given number of peaks, $N_{X_{max}^i}$. As the altitude increases the probability 
of finding a shower with more than one peak goes to zero.

\begin{figure}[!t]
\centering
\includegraphics[width=3.3in]{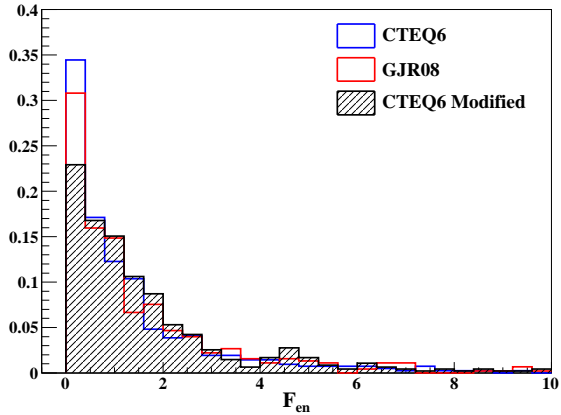}
\caption{Distribution function of parameter $F_{en}$ (see Eq. (\ref{FenDef})) for CTEQ6, GJR08 models and also
for an artificial modification of the CTEQ6 prediction such that the leading particle takes $70\%$ of the neutrino 
energy. Horizontal showers of $10^{20}$ eV injected at sea level at the axis of the FOV are considered.}
\label{Fen}
\end{figure}

The first portion of the cascades is, in general, dominated by the hadronic component. Therefore,
the way in which the primary neutrino energy is distributed can be assessed by,
\begin{equation} 
\label{FenDef}
F_{en} = \frac{\int_0^{X_c} dX\ \frac{dE}{dX} }{ \int_{X_c}^{X_{lim}} dX\ \frac{dE}{dX}},
\end{equation}
where $dE/dX$ is the energy deposition, $X_c=1200$ gcm$^{-2}$ is a characteristic depth that roughly separates 
the hadronic-dominated from the electromagnetic-dominated portions of the shower, and $X_{lim}$ is the maximum 
atmospheric depth reached by the shower. $F_{en}$ is calculated only for those showers that present a first 
maximum at a depth smaller than $X_c$. Fig. \ref{Fen} shows the $F_{en}$ distributions obtained for CTEQ6, GJR08 
and for a modification of the CTEQ6 results in which the leading particle takes 70\% of the neutrino energy, 
corresponding to horizontal $10^{20}$ eV electron neutrinos injected at sea level at the axis of the FOV. It can be 
seen that $F_{en}$ is correlated with the energy taken by the leading particle. The smallest value of the average 
of $F_{en}$ corresponds to CTEQ6 because the leading particle takes about 82\% of the neutrino showers while, in 
the case of GJR08, it takes $\sim 79\%$. The differences between the results obtained for these models are very 
small. In the case in which the energy taken by the leading particle is artificially reduced to $70\%$ (hashed 
histogram), the mean value of $F_{en}$ increases. Note that, in the latter case, the energy extracted from the 
leading particle in order to reduce its average energy, is redistributed among the other daughter particles in 
such a way that the ratio between the energy taken by a given particle and the total energy taken by all 
secondaries, excluding the leading particle, is constant.

The sensitivity of $F_{en}$ to the hadron component of the showers depends on the altitude. Showers injected at sea 
level in regions of high density are dominated by the LPM effect allowing a clearer separation between the hadronic 
and electromagnetic portions of the cascades.

The parameter $F_{en}$ can be very useful to understand neutrino interactions with atmospheric nuclei and, in particular, 
to estimate the energy fraction taken by the leading particle. Any practical application, however, will depend on the 
actual event rate.

\section{Proton and neutrino events}

The interaction length for protons is $\lambda_{pr}(10^{20}\textrm{eV}) \sim 36$ g~cm$^{-2}$ and for neutrinos is 
$\lambda_{\nu}(10^{20}\textrm{eV}) \sim 3.2\times10^7$ g~cm$^{-2}$. The survival probability of an horizontal proton 
that reaches the Earth surface at the vertical axis of the FOV is $\sim \exp(-1000)$, whereas the corresponding probability 
for a neutrino is $\sim \exp(-0.001)$. Therefore, despite the fact that horizontal neutrino and proton showers have very 
different observational characteristics, it is very unlikely to observe a proton interacting so deep in the atmosphere. 

Nevertheless, for a given proton and a neutrino fluxes, there exists a particular slant depth for which the proton and 
neutrino events have the same rate. For any particle type, the probability of interacting in the interval $[X,X+\Delta X]$ 
is given by,
\begin{eqnarray}
P_{int} (E;X,\Delta X) &=& \exp(-X/\lambda(E))\times \nonumber \\ 
&& \left[1-\exp(-\Delta X/\lambda(E)) \right], 
\end{eqnarray}
where $\lambda(E)$ is the interaction length at a given energy. Therefore, solving the equation 
$\phi_{pr}(E)\ P_{int}^{pr}(E;X_0,\Delta X)=\phi_{\nu}(E)\ P_{int}^{\nu}(E;X_0,\Delta X)$, where
$\phi_{pr}(E)$ and $\phi_{\nu}(E)$ are the proton and neutrino fluxes, the slant depth at which
protons and neutrinos can be detected with the same rate is,
\begin{eqnarray}
X(E)&=& \frac{ \lambda_{pr}(E) \lambda_{\nu}(E) }{\lambda_{pr}(E)-\lambda_{\nu}(E)} \times%
\left[ \log\left( \frac{\phi_{\nu}(E)}{\phi_{pr}(E)} \right) + \right.
\nonumber \\
&& \left. \log\left( \frac{1-\exp(-\Delta X/\lambda_{pr}(E))}{%
1-\exp(-\Delta X/\lambda_{\nu}(E))}   \right) \right].
\end{eqnarray}

The function $X(E)$ is obtained by using: $\Delta X = \lambda_{pr}(E)$, the proton-air cross section 
of Sibyll 2.1 \cite{Sibyll2.1}, the neutrino cross section 
$\sigma_{N\nu}^{CC}(E) = 6.04\times10^{-36} (E/\textrm{GeV})^{0.358}$ cm$^2$ \cite{Anchordoqui}, 
the Waxman-Bachall upper limit for the neutrino flux \cite{Waxman} and a power law fit of the Auger 
spectrum \cite{AugerSpec:08}. Therefore, for $E=10^{20}$ eV the event rates for protons and neutrinos 
are of the same order of magnitude for $X\cong142$ gcm$^{-2}$. The latter means that, under the assumptions 
of the present calculation, protons can act as a background for neutrino identification
in inclined events. 
  
\begin{figure}[!t]
\centering
\includegraphics[width=3.3in]{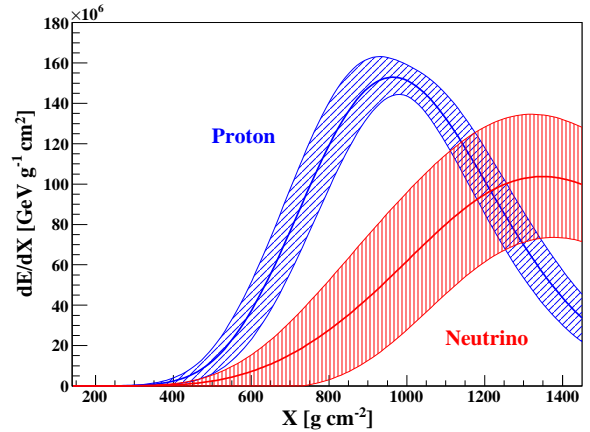}
\caption{Proton and electron neutrino air showers of $10^{20}$ eV, $\theta = 45^{\circ}$ injected at 
142 gcm$^{-2}$.}
\label{Sh45deg}
\end{figure}

Fig. \ref{Sh45deg} shows the average profile and the 68\% CL for proton and electron neutrino induce air
shower longitudinal profiles of zenith angle $\theta=45^\circ$ and primary energy $10^{20}$ eV injected 
at 142 gcm$^{-2}$. Although, the event rate of this kind of showers are similar for both protons and 
neutrinos, the profiles, however, are quite different making a good discrimination possible. More 
specifically, it can be seen that the neutrino showers develop deeper in the atmosphere and present 
larger fluctuations, due to the LPM effect.

\section{Conclusions}

Neutrino detection is of great importance for the understanding of several astrophysical process and, 
in particular, the origin and propagation of the highest energy cosmic rays. JEM-EUSO is technically 
capable of observing neutrino initiated cascades and to discriminate these cascades from hadronic ones. 
Furthermore, the high degree of structuring of the longitudinal neutrino shower profiles opens a spectrum 
of experimental opportunities which JEM-EUSO will be able to explore, with larger or lesser success, 
depending on the exact shape of the neutrino energy spectrum at the highest energies.

\end{document}